\documentclass[preprint,review,12pt]{elsarticle}

\newcommand{\Li}{^6{\rm Li}}
\newcommand{\Be}{^7{\rm Be}}
\newcommand{\bra}{\langle}
\newcommand{\ket}{\rangle}
\def\m{\phantom{-}}

\usepackage{amssymb}
\usepackage{amsmath}
\usepackage{color}

\begin{document}

\begin{frontmatter}


\title{Theoretical calculation of the $p-\,\Li$ radiative capture reaction}


\author[a,b]{Alex Gnech}
\author[c,b]{Laura Elisa Marcucci}

\address[a]{Gran Sasso Science Institute, L'Aquila 67100, Italy}
\address[b]{Istituto Nazionale di Fisica Nucleare sez. di Pisa, Pisa 56127, Italy}
\address[c]{Dipartimento di Fisica, Universit\`a di Pisa, Pisa 56127, Italy}

\begin{abstract}
  We present a new calculation of the $\Li(p,\gamma)\Be$
  radiative capture astrophysical $S$-factor in a cluster model framework.
  We consider several intercluster potentials,
  adjusted to reproduce the $\Be$ bound state properties
  and the $p-\,\Li$ elastic scattering
  phase shifts. Using these potentials, we calculate
  the astrophysical $S$-factor, obtaining
  a good agreement with available data,
  and the photon angular distribution. Finally,
  we discuss the consequences of a hypothetical resonance-like structure
  on the $S$-factor.
\end{abstract}


\end{frontmatter}


\section{Introduction}
\label{sec:intro}

The $\Li$  nucleus  is  not  considered  as  one  of  the  main
Big Bang Nucleosynthesis (BBN)  products,  because  it  is
believed to appear in very small percentages, being a weakly bound nucleus.
However, a measurement of the primordial abundance, using the $\rm{Li}$
absorption line in old halo stars, has revealed an enhancement
compared to standard BBN model predictions~\cite{MA06}.
This is known as the {\it second Lithium problem}, the {\it first} one
being the well known discrepancy between theory and experiment
for the $^7{\rm Li}$ primordial abundance.
A more recent analysis of the data, performed with more sophisticated
models of the stellar atmosphere, seems to reduce this
discrepancy~\cite{RC07,AP09,MS10,KL13}. However, the 
{\it second Lithium problem} has pushed
towards exotic scenarios, as
possible SUSY modification of the BBN model~\cite{MK06}.
In order to exclude or accept these scenarios, it is necessary
to know with high accuracy
the cross sections 
(expressed as astrophysical $S$-factor)
of those reactions that according to BBN contribute
to determine the $\Li$ abundance.
Two of these reactions are the most important: the
$^4{\rm He}(d,\gamma)\Li$ radiative capture,
which creates $\Li$, and the $\Li(p,\gamma)\Be$ radiative capture,
which contributes to destroy $\Li$. The first reaction has been recently
studied in a framework similar to the one proposed here
in Ref.~\cite{Gra2017}. The second reaction
was extensively studied
experimentally~\cite{ZS79,CT87,FC92,RP04,JH13}.
However, large uncertainties in the $S$-factor at the BBN energies
(50--400 keV) remain.
Furthermore, a recent work~\cite{JH13}
has pointed out the possible presence of a resonance
in the BBN energy window, with subsequent suppression at zero energy.
In order to confirm or reject such possibility,
the LUNA Collaboration has also performed
a new campaign of measurements in the Spring of 2018.

The extrapolation of the astrophysical $S$-factor at zero-energy
has been performed 
within the R-matrix approach in Refs.~\cite{SI16,ZL18}, including
somewhat by hand 
the resonance-like structure proposed in Ref.~\cite{JH13}.
On the other hand, all theoretical calculations performed within
the cluster model framework
do not reproduce the claimed resonance. The most important
theoretical studies
were performed using different approaches,
like a two-body phenomenological potential~\cite{JH10,SD11},
an optical potential~\cite{FB80},
a four-cluster model~\cite{KA02} and  the Gamow shell model~\cite{GD17},
obtaining all quite consistent results with each other.
All these studies, however, are lacking of an estimate of the
theoretical uncertainty, especially that arising from
model dependence. Therefore, we present here a new theoretical study
within a cluster model of the $\Li(p,\gamma)\Be$, using
also a two-body phenomenological potential similar to that
of Ref.~\cite{SD11},
but calculating not only
the astrophysical $S$-factor,
but also the angular distribution
of the emitted photon, for which there are also available
data~\cite{CT87}. This will allow us to further verify the agreement
between this theoretical framework and experiment.
We will also investigate on the possible presence of the
resonance structure as suggested by the data of Ref.~\cite{JH13}.

This work is organized as follows: in Sec.~\ref{sec:theory} we will
present the main ingredients of the calculation, while in Sec.~\ref{sec:res}
we will present and discuss our results. Our conclusions are given in
Sec.~\ref{sec:concl}.

\section{Theoretical formalism}
\label{sec:theory}

The cluster model approach is based on the fact that the two colliding nuclei,
$p$ $(J^\pi=1/2^+)$ and $\Li$ $(J^\pi=1^+)$,
can be considered as structureless particle,
which interacts through an {\it ad hoc} potential.
This is tuned to reproduce the $\Be$ properties and the
elastic scattering phase shifts.
Following Ref.~\cite{SD11}, we consider a $p-\,\Li$ potential of the form
\begin{equation}
  V(r)=-V_0\exp{(-a_0r^2)}\,,
\label{eq:vr}
\end{equation}
where $V_0$ and $a_0$ are two parameters, to be chosen by
reproducing the elastic scattering data. We add also a point-like
Coulomb interaction,
\begin{equation}
  V(r)=\alpha \frac{Z_1Z_2}{r} \ ,
\label{eq:pointCoul}
\end{equation}
where $\alpha=1.439975$ MeV fm. All the other coefficients
entering the two-body Schr\"odinger equation
which is solved in this framework are given for completeness
in Table~\ref{tab:params}.
\begin{table}[h]
\begin{center}
  \begin{tabular}{lc}
    \hline
    $m_p$ & $1.00727647$ u\\
    $m_{\Li}$& $6.01347746$ u\\
    $\hbar c$& $197.3269788$ MeV fm\\
    \hline
\end{tabular}
\caption{ \label{tab:params}
  Values of the parameters used in the Schr\"odinger equation.
Note that we have used 
$1$ u $=931.4940954$ MeV.}
\end{center}
\end{table}
All the results that will follow are
obtained using the Numerov algorithm to solve the Schr\"odinger equation
and then further
tested using the R-matrix method (see Ref.~\cite{PD10} and references therein).

\begin{figure}[ht]
  \begin{center}
    \includegraphics[scale=0.65]{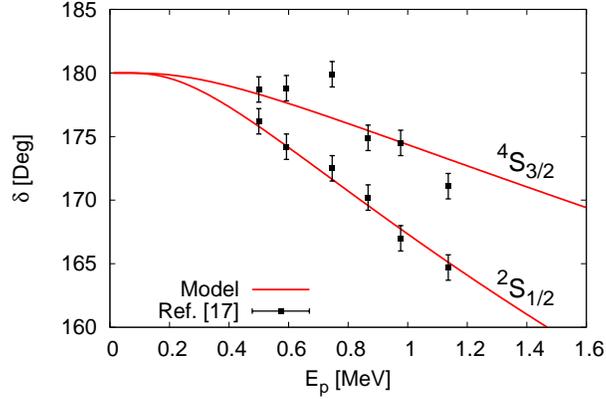}
    \caption{\label{fig:phase-shifts}
      Phase shifts for the ${}^2S_{1/2}$ and the ${}^4S_{3/2}$ partial waves
      as function of the proton energy.
      The data are taken form Ref.~\cite{SD11}. The full red line are 
      the calculated phase shifts with the potential parameters given in
      Table~\ref{tab:chim}.}
  \end{center}
\end{figure}

The parameters of the intercluster potential given in Eq.~(\ref{eq:vr})
are chosen in order to reproduce
the elastic scattering phase shifts, which are derived from partial wave
analysis of the experimental elastic scattering data of Ref.~\cite{SD11}.
In Table~\ref{tab:par-wave}
we report all possible partial waves
up to orbital angular momentum $L=2$ that need to be considered,
both for the doublet $S=1/2$ and quartet $S=3/2$ states, $S$
being the sum of the proton and $\Li$ spins, 1/2 and 1 respectively.
\begin{table}[h]
\begin{center}
\begin{tabular}{l|cc}
& $S=1/2$ & $S=3/2$\\
\hline
$L=0$ & ${}^2S_{1/2}$ & ${}^4S_{3/2}$ \\
$L=1$ & ${}^2P_{1/2}$ ${}^2P_{3/2}$ & ${}^4P_{1/2}$ ${}^4P_{3/2}$ ${}^4P_{5/2}$\\
$L=2$ & ${}^2D_{3/2}$ ${}^2D_{5/2}$ & ${}^4D_{1/2}$ ${}^4D_{3/2}$ ${}^4D_{5/2}$
${}^4D_{7/2}$\\
\hline
\end{tabular}
\caption{ \label{tab:par-wave}
  Partial waves of the $p(J^\pi=1/2^+)-\, \Li(J^\pi=1^+)$ system up to $L=2$.
We indicate with $S$ the total spin.}
\end{center}
\end{table}
While the value of $a_0$ has been fixed and kept as in Ref.~\cite{SD11},
the values of $V_0$ has been obtained minimizing the $\chi^2$ function,
defined as
\begin{equation}
  \chi^2=\sum_i\frac{\left(\delta^i_{\rm{EXP}}(E)
    -\delta^i_{\rm{TH}}(V_0,E)\right)^2}
      {(\Delta\delta^i_{\rm{EXP}})^2}\ .
\label{eq:chi2}
\end{equation}
Here $\delta^i_{\rm{EXP}}(E)$ are the experimental phase shifts
and $\delta^i_{\rm{TH}}(V_0,E)$ are the calculated ones.
The minimization has been performed using the COBYLA algorithm~\cite{MP98}.
The values of $V_0$ and $a_0$ for the various partial waves and the
corresponding $\chi^2$/datum are listed in
Table~\ref{tab:chim}. To be noticed that the phase shift for the ${}^2P$ wave
is given by $\delta_{{}^2P}=\delta_{{}^2P_{1/2}}+\delta_{{}^2P_{3/2}}$
as defined in Ref.~\cite{SD11}.
In Fig.~\ref{fig:phase-shifts}
we report the
experimental values and the calculated phase shifts for the $S$ waves.
As we can see from the figure, a nice agreement is found for the
$S$-wave phase shifts, especially for the ${}^2S_{1/2}$.
\begin{table}[ht]
  \begin{center}
    \begin{tabular}{l|ccc}
      \hline 
        wave &  $V_0$ (MeV) & $a_0$ (fm$^{-2}$)& $\chi^2/$datum\\ 
        \hline
        ${}^2S_{1/2}$ & 124.63 & 0.15 & 0.4\\
        ${}^4S_{3/2}$ & 141.72 & 0.15 & 3.6\\
        ${}^2P$ & 67.44  & 0.1 & 1.9\\
        \hline
    \end{tabular} 
    \caption{Values for the parameters of the
      Gaussian potential and $\chi^2/$datum for the different
      partial waves.}\label{tab:chim}
  \end{center}
\end{table}

The $p-\,\Li$ potential of the form of Eq.~(\ref{eq:vr}) is used
also in order to describe the $\Be$ nucleus. In this case
we need to reproduce the binding energies of the two bound states,
the ground state (GS) $J^\pi=3/2^-$ with $E=-5.6068$ MeV
and the first excited state (FES) $J^\pi=1/2^-$ with $E=-5.1767$
MeV~\cite{DT02}.
We fixed again the parameter $a_0$ as in Ref.~\cite{SD11},
while in order to obtain $V_0$ we impose 
that the calculated binding energies reproduce the experimental ones up to the
sixth digit. Moreover, we have evaluated also the
asymptotic normalization coefficient
(ANC), defined as
\begin{equation}
    {\rm{ANC}}=\frac{u_{LS}(r)}{\sqrt{2k}W_{L+1/2}(2kr,\eta)}\,,
\label{eq:anc}
\end{equation}
where $u_{LS}(r)$ is the radial part of the wave function (see below),
$r$ is the intercluster
distance, $k=\sqrt{\frac{2\mu E}{\hbar^2}}$ with
$\mu=\frac{m_pm_{\Li}}{m_p+m_{\Li}}$, $E$ the energy of the bound state, and
$W_{L+1/2}(2kr,\eta)$
is the Whittaker function~\cite{AS65}, with $\eta$ defined as
\begin{equation}
    \eta=1.439975\times Z_pZ_{\Li}\frac{\mu}{2k\hbar^2}\,.
    \label{etabar}
\end{equation}
In Table~\ref{tab:bspar}
we report the values for $V_0$ and $a_0$, 
and the calculated value of the binding energies and ANCs for both the GS
and FES. Note that, to our knowledge, there are no experimental data
for the ANCs.
\begin{table}[h]
  \begin{center}
    \begin{tabular}{l|cccc}
      \hline 
        $J^\pi$ &  $V_0$ (MeV) & $a_0$ (fm$^{-2}$)& $E$ (MeV) & ANC\\ 
        \hline
        $3/2^-$ & 254.6876510 & 0.25 & -5.606800 & 2.654\\
        $1/2^-$ & 252.7976803 & 0.25 & -5.176700 & 2.528\\
        \hline
    \end{tabular} 
    \caption{Values for the parameters of the
      Gaussian potential of Eq.~(\ref{eq:vr})
      and the calculated binding energy and ANC for 
      both the GS and the FES.
      }\label{tab:bspar}
  \end{center}
\end{table}

Having determined the $\Be$ and $p-\,\Li$ wave functions, we can proceed
to evaluate the radiative capture cross section and angular distribution.
Let us consider the generic reaction $A_1+A_2\rightarrow A_3+\gamma$.
We write the scattering wave function as
\begin{eqnarray}
  \psi_{1,2}(\vec{r},p)&=&\frac{\sqrt{4\pi}}{p}\sum_{LSJJ_z}
  i^L\, \sqrt{2L+1}\,
  \bra J_1 M_{1} J_2 M_{2}| S J_z\ket\nonumber\\
  &&\bra SJ_z L 0| J J_z \ket
  \,\psi_{1,2}^{LSJJ_z}(\vec{r},p)\,,
\label{eq:psi12}
\end{eqnarray}
with
\begin{equation}
  \psi_{1,2}^{LSJJ_z}(\vec{r},p)=
  R_{LSJ}(r,p)\,\Big[Y_L(\hat{r})\otimes\chi_S\Big]_{JJ_z}\,,
  \label{eq:psi12_LSJ}
\end{equation}
where $p$ is the relative momentum of the two particles, $\vec{r}$
the intercluster distance, $L$, $S$ and $J$ the total orbital,
spin and angular momentum of the two nuclei,
with $J_1,M_1$ and $J_2,M_2$ being the total
angular momenta and third components of the two nuclei.
The function $R_{LSJ}(r,p)$ is the scattering wave function,
that has been
determined solving the two-body Schr\"odinger equation similarly to
what done in Ref.~\cite{Gra2017}.
For the bound states of the final nucleus $A_3$ we write the wave function
as
\begin{equation}
  \psi_{3}^{J_3M_{3}}(\vec{r})=
  u_{L_3S_3}(r)\,\Big[Y_{L_3}(\hat{r})\otimes\chi_{S_3}\Big]_{J_3M_{3}}\,,
  \label{eq:bound}
\end{equation}
where $\vec{r}$ is again the intercluster distance.
The function $u_{L_3S_3}(r)$ has also been determined as
explained above.
The total cross section for a radiative capture in a bound state with
total angular momentum $J_3$ is written as
\begin{eqnarray}
  \sigma_{J_3}(E)&=&\frac{32\pi^2}{(2J_1+1)(2J_2+1)}\,\frac{\alpha}{v_{\rm rel}}
  \,\frac{q}{1+q/m_3}\nonumber\\
  &&\qquad\times\sum_{\Lambda\geq1}\sum_{LSJ}
  \Big(|E_{\Lambda}^{LSJ,J_3}|^2+|M_{\Lambda}^{LSJ,J_3}|^2\Big)
  \label{eq:sigtot}\,,
\end{eqnarray}
where $\alpha=e^2/4\pi$, $v_{\rm rel}$ is the relative velocity of the
two incoming particles, $q$ is the photon momentum and $m_3$ is the mass of
$A_3$ nucleus. Finally, $T_{\Lambda}^{LSJ,J_3}$, with $T=E/M$,
are the reduced matrix element of the electromagnetic
operator and $\Lambda$ is the multipole order.
Using the Wigner-Eckart theorem, they are defined as
\begin{equation}
  T^{LSJ,J_3}_\Lambda=\bra \psi_{1,2}^{LSJJ_z}(\vec{r},p)|
  T_{\Lambda\lambda}|\psi_{3}^{J_3M_{3}}(\vec{r})\ket\frac{\sqrt{2J_3+1}}
  {\bra J_3 M_{3} \Lambda \lambda | J J_z\ket}\,,
  \label{eq:tlambda}
\end{equation}
where $\lambda=\pm1$ is
the photon polarization.
In our calculation we include only the electric operator,
which is typically larger than the magnetic one. Then, in 
the long-wavelength approximation~\cite{JD95}, by using
Eqs.~(\ref{eq:psi12_LSJ}) and~(\ref{eq:bound}), it results
\begin{eqnarray}
  E_{\Lambda}^{LSJ,J_3}&=&(-1)^{2J_f+\Lambda+L+S-J}
  \hat{J}\hat{J_3}\hat{L_3}\hat{\Lambda}
  \,\bra L_3 0 \Lambda 0 | L 0\ket 
  \nonumber\\
  &&\times\begin{Bmatrix}
      J   & L   & S \\
      L_3 & J_3 &\Lambda
    \end{Bmatrix}
  \frac{Z_e^{(\Lambda)}}{(2\Lambda+1)!!}
  \sqrt{\frac{\Lambda+1}{\Lambda}}\frac{q^\Lambda}{\sqrt{4\pi}p}\nonumber\\
  &&\times\int_0^\infty dr\, r^2\,u_{L_3S_3}(r)r^\Lambda
  R_{LSJ}(r,p)\delta_{S,S_3}\label{eq:eleop}
  \,.
\end{eqnarray}
Here we have defined $\hat{x}=\sqrt{2x+1}$ and
\begin{equation}
  Z_e^{(\Lambda)}=Z_1\left(\frac{m_2}{m_1+m_2}\right)^\Lambda+
  Z_2\left(-\frac{m_1}{m_1+m_2}\right)^\Lambda\,
  \label{eq:zeff}
\end{equation}
is the effective charge, in which $Z_1(Z_2)$ is the charge
and $m_1(m_2)$ is the mass of the $A_1(A_2)$ nucleus.
Given the radial wave functions $u_{L_3S_3}(r)$
and $R_{LSJ}(r,p)$, the one-dimensional integral of
Eq.~(\ref{eq:eleop}) is simple and performed with standard
numerical techniques.
The astrophysical $S$-factor is then defined as
\begin{equation}
  S_{J_3}(E)=E\exp(2\pi\eta)\sigma_{J_3}(E)\,,
\label{eq:sfactor}
\end{equation}
where $\sigma_{J_3}(E)$ is the total cross section of
Eq.~(\ref{eq:sigtot}) and
$\eta$ is defined in Eq.~(\ref{etabar}). 

The other observable of interest is the
photon angular distribution, which can be written as
\begin{equation}
  \sigma_{J_3}(E,\theta)=\sigma_0(E)\sum_{k}a_k^{J_3}(E) P_k(\cos\theta)\,,
  \label{eq:sigma_theta}
\end{equation}
where $\sigma_0(E)$ is a kinematic factor defined as
\begin{equation}
  \sigma_0(E)=\frac{16\pi^2}{(2J_1+1)(2J_2+1)}\frac{\alpha}{v_{\rm rel}}
  \frac{q}{1+q/m_3}\,,\label{eq:sig0}
\end{equation}
and $P_k(\cos\theta)$ are the Legendre polynomials. The coefficients
$a_k$ are given by
\begin{eqnarray}
  a_k^{J_3}(E)&=&\sum_{LL'SJJ'\Lambda\Lambda'}(-)^{J+J'+J_3+L'+\Lambda+S+1}
  \,i^{L+L'+\Lambda+\Lambda'}\nonumber\\
  &&\times\hat{L}\hat{L'}\hat{\Lambda}\hat{\Lambda'}
  \hat{J}\hat{J'}\,\bra L 0 L' 0|k 0\ket   \begin{Bmatrix}
    L   & L'   & k \\
    J' & J & S
  \end{Bmatrix}\nonumber\\
  &&\times\begin{Bmatrix}
      J'   & J   & k \\
      \Lambda & \Lambda' & J_3
    \end{Bmatrix}
  \sum_{\lambda=\pm1}\bra \Lambda' -\lambda \Lambda \lambda |k 0\ket
  \nonumber\\
  &&\times\left(\lambda M_{\Lambda'}^{L'SJ',J_3}+E_{\Lambda'}^{L'SJ',J_3}\right)
  \left(\lambda M_{\Lambda}^{LSJ,J_3}+E_{\Lambda}^{LSJ,J_3}\right)\,.
  \label{eq:ak}
\end{eqnarray}
The photon angular distribution can be casted in the final form
\begin{equation}
  \sigma_{J_3}(E,\theta)
  =\sigma_{J_3}(E)\left(1+\sum_{k\geq1}A^{J_3}_k(E) P_k(\cos\theta)\right)\,,
  \label{eq:angulardistr}
\end{equation}
where $\sigma_{J_3}(E)$ is defined in Eq.~(\ref{eq:sigtot}),
and $A^{J_3}_k(E)=a_k^{J_3}(E)/a_0^{J_3}(E)$.

\section{Results}
\label{sec:res}

In this section we compare our theoretical predictions for the
astrophysical $S$-factor and the
angular distribution of the emitted photon
with the available experimental data. In the last subsection,
we also discuss the possibility of introducing in our model the resonance
proposed in Ref.~\cite{JH13}.

Before discussing the results, we note that in the $p-\,\Li$ reaction
the open 
${}^3{\rm He}-\,^4{\rm He}$ channel should in principle
be included. However, we do not consider
this channel in our work. This can be done because
the experimental phase shifts of Ref.~\cite{SD11} used to fit
our potential were obtained considering only the $p-\,\Li$ channel.
Therefore the ${}^3{\rm He}-\,^4{\rm He}$ channel
results to be hidden in the experimental phase shifts that we reproduce
with our potential. On the other hand, for the $\Be$ bound states,
the ${}^3{\rm He}-\,^4{\rm He}$
component needs to be considered, and this is
done phenomenologically, introducing in our calculation the
spectroscopic factors, as explained in the next subsection.

\subsection{The Astrophysical $S$-factor}
\label{subsec:sfactor}

The main contribution to the radiative capture reaction 
$\Li(p,\gamma)\Be$ cross section (and therefore
astrophysical $S$-factor) comes from the electric dipole ($E1$) transition.
 The structure of the electric operator in the long wavelength
  approximation implies a series of selection rules due to the presence of
  the Wigner-6j coefficient as shown in
Eq.~(\ref{eq:eleop}).
Therefore, the only waves allowed by the $E1$ transition operator
up to $L=2$ are ${}^2S_{1/2}$, ${}^2D_{3/2}$ and ${}^2D_{5/2}$
for the GS, and  
${}^2S_{1/2}$ and ${}^2D_{3/2}$ for the FES.
To evaluate the ${}^2D$ waves, we use the same potential used for the
${}^2S_{1/2}$ wave, but we have changed only the angular momentum
$L$ in the Schr\"odinger equation and we have 
imposed that the waves ${}^2D_{3/2}$ and ${}^2D_{5/2}$
are identical in the radial part.
From the calculation, it turns out that up to energies
of about 400 keV, the contribution
of the ${}^2D$ waves is very small. However, for higher values of the energy,
this contribution becomes significant.

\begin{figure}[t]
  \begin{center}
    \includegraphics[scale=0.65]{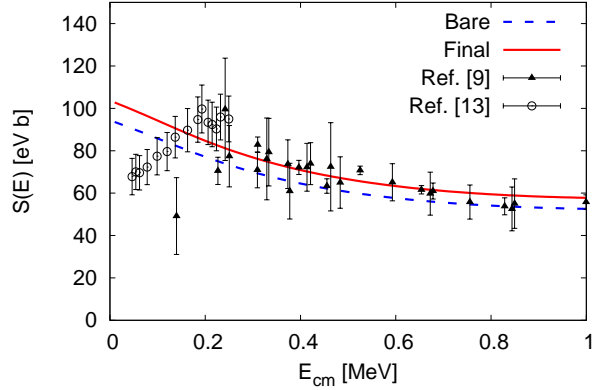}
    \caption{\label{fig:sftot}
      Total astrophysical
      $S$-factor for the $\Li(p,\gamma)\Be$ radiative capture reaction.
      The (blue) dashed line is the bare calculation, while
      the (red) full line is the obtained including
      the spectroscopic factors ${\cal{S}}_0$ and ${\cal S}_1$
      of Table~\ref{tab:spectro}.
      The data are taken from
      Refs.~\cite{ZS79} and~\cite{JH13}.}
  \end{center}
\end{figure}

In Fig.~\ref{fig:sftot} we compare our results for the
astrophysical $S$-factor with the experimental
data of Ref.~\cite{ZS79} and~\cite{JH13}. The calculation is
performed summing up the
contributions to both the GS and the FES.
Since the data of Ref.~\cite{JH13} are still under debate, 
in discussing the results of Fig.~\ref{fig:sftot} we will
consider only the data of Ref.~\cite{ZS79}. By inspection of the
figure, we can conclude that our calculated (bare) $S$-factor is
systematically lower than the data.
The reason can be simply traced back
to the fact that in our model we do not take into account
the internal structure of $\Li$ and $\Be$.
In order to overcome this limitation,
we introduce the spectroscopic factor $\cal{S}$,
for both bound states of $\Be$,
so that the total cross section can be rewritten as
\begin{equation}
  \sigma(E)={\cal{S}}_0^2\sigma^{\rm{bare}}_0(E)+
        {\cal{S}}_1^2\sigma^{\rm{bare}}_1(E)\,.
\end{equation}
Here $\sigma^{\rm{bare}}_0(E)\,(\sigma^{\rm{bare}}_1(E))$ 
and ${\cal{S}}_0({\cal{S}}_1)$ are the calculated bare cross section
and spectroscopic factor
for the transition to the  GS (FES) of $\Be$.

In order to determine the two spectroscopic factors ${\cal{S}}_0$
and ${\cal{S}}_1$, we proceed as follows: we notice that
in Ref.~\cite{ZS79} there are two sets of data,
which corresponds to the radiative
capture to GS and FES, and the total $S$-factor
is given by multiplying the
data for the relative branching ratio (BR).
Therefore, we divide the two data sets for the corresponding BR and
we fit the spectroscopic factors, calculating the $S$-factor for GS and FES
captures separately.
In such a way we are able to reproduce not only the total $S$-factor but also the
experimental BR for the FES radiative capture of $\sim39\%$~\cite{ZS79},
defined as ${\cal{S}}_1^2\sigma^{\rm{bare}}_1(E)/\sigma(E)$.
The values of the spectroscopic
factors and the $\chi^2/$datum defined according to
Eq.~(\ref{eq:chi2}), using the data of Ref.~\cite{ZS79}, before
($\chi^2_0/$datum) and after ($\chi^2_{\cal S}/$datum) adding
the spectroscopic factors are given in Table~\ref{tab:spectro}.
From the values of the $\chi^2_0/$datum given
in Table~\ref{tab:spectro}, it is possible to conclude that the 
description of the radiative capture reaction to the GS
using the bare wave function 
is quite accurate, while this is not the case for the FES.
\begin{table}[h]
  \begin{center}
    \begin{tabular}{lccc}
      \hline 
      $J^\pi$ & ${\cal S}$ & $\chi_0^2/$datum & $\chi^2_{\cal_S}/$datum\\ 
      \hline
      $3/2^-$ & 1.003 & 0.064  & 0.064 \\
      $1/2^-$ & 1.131 & 2.096  & 0.219 \\
      \hline
    \end{tabular} 
    \caption{Spectroscopic
      factors ${\cal S}$ and $\chi^2/$datum obtained by fitting
      the data of Ref.~\cite{ZS79}, before
      ($\chi^2_0/$datum) and after ($\chi^2_{\cal S}/$datum) adding
      the spectroscopic factors themselves. With $J^\pi=3/2^-$ and $J^\pi=1/2^-$
      we indicate the GS and FES of $\Be$.}\label{tab:spectro}
  \end{center}
\end{table}

In order to extrapolate the astrophysical $S$-factor at zero energy, we
perform a polynomial fit of our
calculated points up to second order, i.e. we rewrite the $S$-factor $S(E)$
in the energy range between $0$ and $300$ keV as
\begin{equation}
  S(E)=S(0)+S_1(0)E+S_2(0)E^2\,.\label{eq:sfexp}
\end{equation}
In Table~\ref{tab:fit} we report
the values obtained for $S(0)$, $S_1(0)$, and $S_2(0)$
in the cases of the GS, FES and the total GS+FES captures.
\begin{table}[h]
  \begin{center}
    \begin{tabular}{c|ccc|ccc}
      & GS  & FES & GS+FES & Ref.~\cite{JH10} & Ref.~\cite{SD11} & Ref.~\cite{FB80} \\ 
      \hline
      $S(0)$ [eV b]          & $\m63.2$& $\m40.7$& $\m103.9$&$\m98.5$&$\m106$&$\m108$\\
      $S_1(0)$ [eV b/MeV]    & $ -63.9$& $ -41.2$& $ -105.1$&$ -71.5$&$ -215$&$ -130$\\
      $S_2(0)$ [eV b/MeV$^2$]& $\m27.3$& $\m17.7$& $\m45.0$ &$\m32.5$&$\m312$&$\m81.7$ \\
      \hline 
    \end{tabular} 
    \caption{Expansion coefficients for the polynomial fit of the $S$-factor
      as defined in Eq.~(\ref{eq:sfexp}). For our work we report the
      expansion for the GS, the FES, and the sum of the two. For Refs.~\cite{JH10,SD11,FB80}
      we report the expansion obtained from the fit of digitalized curves of the
      total S-factor.}
    \label{tab:fit}
  \end{center}
\end{table}
The results of the table can be compared with those obtained with other
phenomenological models
in Refs.~\cite{JH10,SD11,FB80}.
In particular, we can conclude that 
our results for $S(0)$ is within $3\%$ 
compared to Refs.~\cite{JH10,SD11,FB80}.
As regarding the shape
of the $S$-factor, determined by $S_1(0)$ and $S_2(0)$, our results
are quite in agreement with those of Ref.~\cite{JH10} and~\cite{FB80}.
On the other hand, the results obtained in Ref.~\cite{SD11}
with an approach similar to ours, give a higher value for $S_1(0)$ and $S_2(0)$.
The origin of this discrepancy is still unknown.
The results of Ref.~\cite{KA02}, although obtained with
a more sophisticated model than the one presented here, are consistent
with ours, while those of Ref.~\cite{GD17}
show a different energy dependence.
All the theoretical calculations, except the studies
of Refs.~\cite{FC92,RP04},
agree in a negative slope in the $S$-factor at low energies,
and none of them predict a resonance structure, as suggested instead
by the data of Ref.~\cite{JH13}.

\begin{figure}[h]
  \begin{center}
    \includegraphics[scale=0.65]{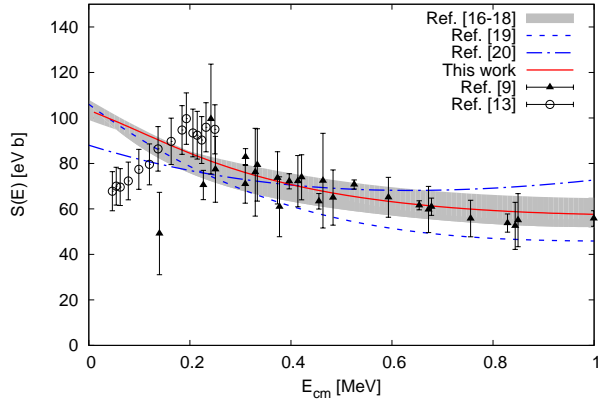}
    \caption{\label{fig:sfcomparison}
      Comparison between our predictions
      (full red line) and the studies available in the
      literature. We show within a (gray) band
      the calculated astrophysical $S$-factors of
      Refs.~\cite{JH10,SD11,FB80}, obtained with
      phenomenological potentials.
      For completeness we also report
      the results of Ref.~\cite{KA02} (blue dashed line) and of
      Ref.~\cite{GD17} (blue dot-dashed line).}
  \end{center}
\end{figure}

In order to estimate the theoretical uncertainty
arising from a calculation performed in the phenomenological
two-body cluster approach, we have reported in Fig.~\ref{fig:sfcomparison}
within a (gray) band all the results
available in the literature.
As we can conclude
by inspection of the figure, the theoretical error
which can be estimated by the band is quite
significant,
but of the same order of
the experimental errors on the data.
If we take into account all the results obtained
with the phenomenological potentials of Refs.~\cite{JH10,SD11,FB80}
and the results of the present work, 
we obtain for the $S$-factor at zero energy the value
\begin{equation}
  S(0)= (103.5\pm 4.5)\ {\rm eV\ b}\,.
  \label{eq:s0}
\end{equation}
We remark that also the value for $S(0)$ obtained in Ref.~\cite{PD10}
is within this range.
  
\subsection{Angular distribution of photons}
\label{subsec:angular}

We present in this section the photon angular
distribution results obtained within the framework
outlined in Sec.~\ref{sec:theory}, and we compare our results
with the data of Ref.~\cite{CT87}.
This provides a further check on our model.

By using Eq.~(\ref{eq:angulardistr}),
we have found that the main contribution to the $A_k^{J_3}(E)$ coefficients
comes from the interference of the
$E1$ operator generated by the ${}^2S_{1/2}$ wave
with the $E1$ operator generated
by the ${}^2D$ waves and with the $E2$ operator generated by the ${}^2P$ waves.
Note that for the ${}^2P$ and ${}^2D$ waves, we do not have a complete
set of data for the phase shifts in all the possible
total angular momentum $J^\pi$. 
Therefore we use the same radial function for
the ${}^2D_{3/2}$ and ${}^2D_{5/2}$ waves,
and also for the ${}^2P_{1/2}$ and ${}^2P_{3/2}$ waves.
The relative phases for these waves, being arbitrary, are fixed in order to have the best
description of the data of Ref.~\cite{CT87}.
   
\begin{figure}[h]
  \begin{center}
    \includegraphics[scale=0.65]{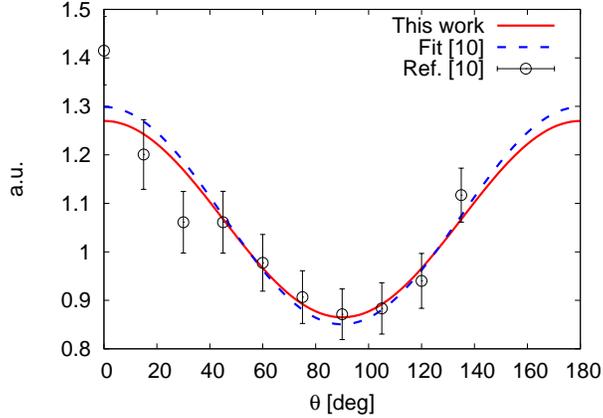}
    \caption{\label{fig:angdist0}
      Photon angular distribution for the radiative capture to the
      GS for $E_p=0.5$ MeV. Our calculation (full red line) is compared
      to the fit (dashed blue line) and the data of Ref.~\cite{CT87}.}
  \end{center}
\end{figure}

\begin{figure}[h]
  \begin{center}
    \includegraphics[scale=0.65]{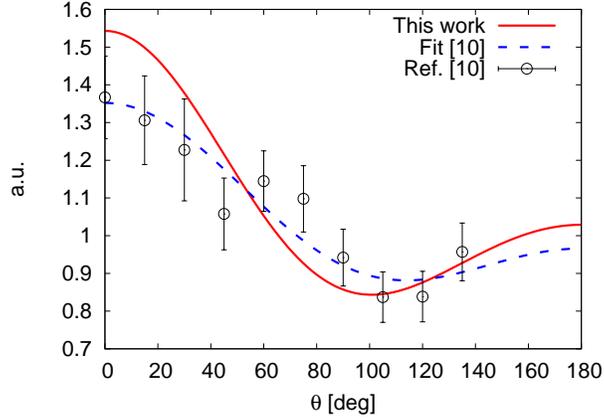}
    \caption{\label{fig:angdist1}
      The same as Fig.~\ref{fig:angdist0} for the FES.}
  \end{center}
\end{figure}

The results for the $A_k^{J_3}$ coefficients for various incident
proton energies $(E_p)$ are reported in
Tables~\ref{tab:t32} and~\ref{tab:t12}, where
they are compared with the values fitted on the
experimental data of Ref.~\cite{CT87}.
\begin{table}[h]
  \begin{center}
    \begin{tabular}{c|cc}
      \hline 
      $k\,(J_3=3/2)$ & This work & Fit of Ref.~\cite{CT87}\\ 
      \hline
      \multicolumn{3}{c}{$E_p=500$ keV}\\
      \hline
      1 &0.000   &   -   \\
      2 &0.270   & $0.299\pm0.045$\\
      3 &0.000   &   -  \\
      \hline
      $\chi^2/$datum & 0.95 & 0.90 \\
      \hline 
      \multicolumn{3}{c}{$E_p=800$ keV}\\
      \hline
      1 &0.000   &   -    \\
      2 &0.375   & $0.390\pm0.031$      \\
      3 &0.000   &   -    \\
      \hline
      $\chi^2/$datum & 0.79 & 1.17 \\
      \hline 
      \multicolumn{3}{c}{$E_p=1000$ keV}\\
      \hline
      1 &0.000   &   -   \\
      2 &0.422   & $0.368\pm0.036$     \\
      3 &0.000   &   -   \\
      \hline
      $\chi^2/$datum & 1.61 & 1.21 \\
      \hline 
    \end{tabular} 
    \caption{\label{tab:t32}
      Values of the coefficients $A_k^{3/2}(E)$ for three proton energies
      compared with the fit to the data of Ref.~\cite{CT87}.
    The $\chi^2/$datum is also reported.}
  \end{center}
\end{table}
\begin{table}[h]
  \begin{center}
    \begin{tabular}{c|cc}
      \hline 
      $k\,(J_3=1/2)$ & This work & Fit of Ref.~\cite{CT87} \\ 
      \hline
      \multicolumn{3}{c}{$E_p=500$ keV}\\
      \hline
      1 &0.214   &$0.193\pm0.055$      \\
      2 &0.286   &$0.159\pm0.074$      \\
      3 &0.043   &      -              \\
      \hline
      $\chi^2/$datum & 1.71 & 0.78 \\
      \hline 
      \multicolumn{3}{c}{$E_p=800$ keV}\\
      \hline
      1 &0.263   &$0.283\pm0.042$      \\
      2 &0.398   &$0.257\pm0.051$      \\
      3 &0.085   &     -               \\
      \hline
      $\chi^2/$datum & 3.12 & 0.76 \\
      \hline 
      \multicolumn{3}{c}{$E_p=1000$ keV}\\
      \hline
      1 &0.280   &$0.205\pm0.043$      \\
      2 &0.448   &$0.281\pm0.054$      \\
      3 &0.115   &     -               \\
      \hline
      $\chi^2/$datum & 6.10 & 1.67 \\
      \hline 
    \end{tabular} 
    \caption{\label{tab:t12}
      Same as Table~\ref{tab:t32}, but for the coefficients $A_k^{1/2}(E)$.}
  \end{center}
\end{table}
In Figs.~\ref{fig:angdist0} and~\ref{fig:angdist1}
we report the calculated angular distribution of the emitted photon for
the capture to the GS  and to the FES, respectively of $E_p=0.5$ MeV. 
The data of Ref.~\cite{CT87} are also shown.
The theoretical values are in nice agreement with the fitted data for the GS.
In particular, the $A_1^{3/2}$ coefficient,
obtained using Eq.~(\ref{eq:ak}), results to be
\begin{equation}
  A_1^{3/2}\propto E_1^{0\frac{1}{2}\frac{1}{2},\frac{3}{2}}
  \left(E_2^{1\frac{1}{2}\frac{1}{2},\frac{3}{2}}
  -E_2^{1\frac{1}{2}\frac{3}{2},\frac{3}{2}}\right)+\ldots \ ,
  \label{eq:a1-32}
\end{equation}
where the dots indicate the interference's between the $E1$ generated by
the $D$ waves, which
give a negligible contribution.
If now we suppose that
$E_2^{1\frac{1}{2}\frac{1}{2},\frac{3}{2}}\simeq E_2^{1\frac{1}{2}\frac{3}{2},\frac{3}{2}}$,
the value for $A_1^{3/2}$ goes to zero, explaining why we
do not need this coefficient to reproduce the data.
In our case the values of the coefficients $A_1^{3/2}$
are exactly zero because we use the same radial function for
different $J^\pi$. The same happens also for $A_3^{3/2}$.
As regarding to the capture to the FES, our calculation shows
some disagreements compared to the values obtained by fit to the data
of Ref.~\cite{CT87}, although these are affected by significant
uncertainties.
In this case, in fact, there is no cancellation as in Eq.~(\ref{eq:a1-32}), and
therefore the values of the
coefficients $A_k^{1/2}$ are strongly dependent on the ${}^2P$ and ${}^2D$
waves, which are very uncertain.
For this reason the disagreement
with the data can be considered acceptable.

\begin{figure}[h]
  \begin{center}
    \includegraphics[scale=0.65]{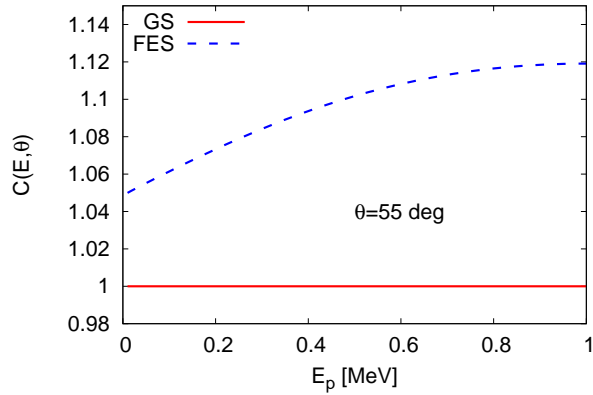}
    \caption{\label{fig:ce}
      Correction coefficient $C(E,\theta)$ defined in Eq.~(\ref{eq:cea})
      for the GS (full red line) and
      for the FES (dashed blue line) capture at $\theta=55^\circ$.}
  \end{center}
\end{figure}

The photon angular distribution has a noticeable impact on the experimental
measurements of the $S$-factor. Many experiments are done measuring
the photon emitted at a fixed angle ($\theta$)
respect to the beam
axis. Therefore the measured cross section must
be corrected by a factor related to the angular distribution.
We take into account this effect writing 
the total cross section as
\begin{eqnarray}
  \sigma(E)=\frac{2\pi}{\phi_2-\phi_1}\,\,\frac{
    \sigma_{exp}(E;\theta_1\theta_2;\phi_1\phi_2)}
        {\int_{\theta_1}^{\theta_2}d\theta\,\sin\theta\, C(E,\theta)}\,,
\label{eq:sigmae}
\end{eqnarray}
where $\sigma_{exp}(E;\theta_1\theta_2;\phi_1\phi_2)$
is the measured cross section
integrated over the solid angle covered by the detector and
\begin{eqnarray}
  C(E,\theta)=1+\sum_{k\geq1}A_k(E)P_k(\cos\theta)\,.
\label{eq:cea}
\end{eqnarray}
In Eq.~(\ref{eq:sigmae}) we call with $\phi$ the other polar angle.
Many experiments put the detector at $\theta\simeq 55^\circ$,
since $P_k(\cos 55^\circ)\simeq0$ and therefore
the contribution of $A_2(E)$ can be neglected.
However, the contribution from the
$A_1(E)$ coefficient can not be always neglected.
Using our calculation, we can estimate the impact of the angular distribution
of the photon to the measurement of the $\Li(p,\gamma)\Be$ reaction. 
In Fig.~\ref{fig:ce} we evaluated the coefficients $C(E,\theta=55^\circ)$
for the capture to the  GS and the FES, neglecting the
physical dimension of the detector. By inspection of the figure
we can conclude that the correction given by the photon
angular distribution is negligible for capture in the GS. This can
be traced back to the fact that the
coefficient $A_1^{J_3}\simeq0$. However, 
$A_1^{J_3}\simeq6-9\%$ for the FES in the region of interest of the BBN.
This can have consequences for the different experimental determinations,
affecting their systematic error estimate.
% 
%  Before concluding the discussion we want to remark that the poor knowledge on
%  the phase shifts on the $P$ and $D$ waves introduce a lot of uncertainties we
%  cannot control. Therefore the predictions reported in this sections suffer of
%  big systematic uncertainties.

\subsection{The ``He''-resonance}
\label{subsec:heres}

In a recent work~\cite{JH13}, He {\it et al.} considered the possibility
of introducing a resonance-like structure
in the $\Li(p,\gamma)\Be$ $S$-factor data at low energies, and they estimated
the energy and width in the proton decay channel to
be $E_R=195$ keV and $\Gamma_p=50$ keV, respectively. The total
angular momentum of the resonance
reported in Ref.~\cite{JH13} can be either
$J^\pi=1/2^+$ or $J^\pi=3/2^+$. In this section
we give for granted the existence
of this resonance, and we
explore the effects of introducing such a resonance
structure in our model. The comparison with the available data
will tell us whether this assumption is valid or not.

\begin{figure}[h]
  \begin{center}
    \includegraphics[scale=0.65]{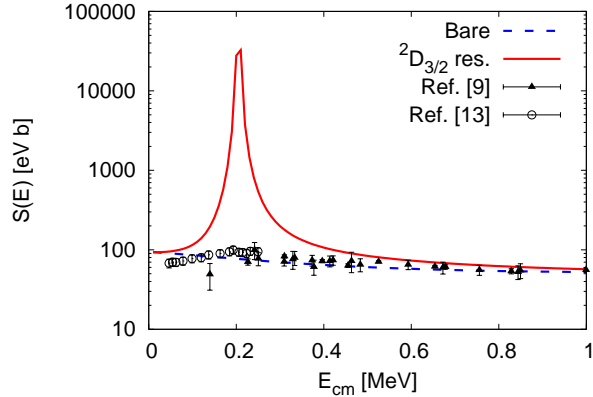}
    \caption{\label{fig:sfresd}
      Bare astrophysical $S$-factor (blue dot-dashed line)
      to which it is summed a resonance structure in
      the ${}^2D_{3/2}$ wave (full red line).
      See text for more details.}
  \end{center}
\end{figure}

The first step of our study consists in constructing
the nuclear potentials in such a way that we obtain 
$190$ keV $<E_R<$ $200$ keV and we reproduce
the width of the resonance in the $S$-factor data.
In a first calculation, we consider to introduce the resonance in the
partial wave of spin 1/2. In particular we use the wave ${}^2S_{1/2}$
for $J^\pi=1/2^+$ and ${}^2D_{3/2}$ for $J^\pi=3/2^+$.
In both the cases, we were not able to find parameters $V_0$ and $a_0$
(see Eq.~(\ref{eq:vr})) that give a consistent description of all the available data. 
For the ${}^2S_{1/2}$ the introduction of such a resonance is completely
inconsistent with the experimental phase shifts.
For the ${}^2D_{3/2}$ we do not have experimental constrains on
the experimental phase shifts, but we were not able to obtain
the strength of the resonance as given in the data of Ref.~\cite{JH13}. 
The best result obtained adding the resonance in the ${}^2D_{3/2}$ wave is given
in Fig.~\ref{fig:sfresd}.

\begin{figure}[h]
  \begin{center}
    \includegraphics[scale=0.65]{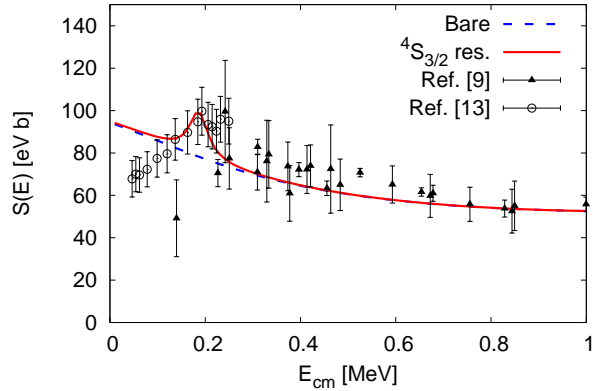}
    \caption{\label{fig:sfres}
      Bare astrophysical $S$-factor (blue dot-dashed line)
      to which it is summed a resonance structure in
      the ${}^4S_{3/2}$ wave (full red line). See text for more details.}
  \end{center}
\end{figure}

\begin{figure}[h]
  \begin{center}
    \includegraphics[scale=0.65]{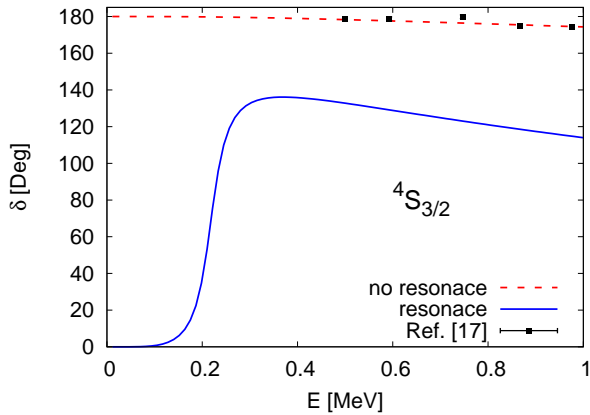}
    \caption{\label{fig:phsres}
      Elastic scattering phase shifts for the ${}^4S_{3/2}$ wave
      calculated in the
      case of resonance (blue full line) and no resonance (red dashed line),
      are compared with the data of Ref.~\cite{SD11}.}
  \end{center}
\end{figure}

In a second calculation, we considered the GS of $\Be$ to be a mixed
state of spin $1/2$ and $3/2$. In this way the $E1$ operator can couple
the scattering wave ${}^4S_{3/2}$ to the ${}^4P_{3/2}$ component of the GS. Therefore,
we can introduce the $J^\pi=3/2^+$ resonance in the ${}^4S_{3/2}$ partial
wave. In this calculation, we use the ${}^2P_{3/2}$ radial wave function for
the ${}^4P_{3/2}$ component of the GS.
We select as potential parameters for the
${}^4S_{3/2}$ component
$V_0=438.7$ MeV and
$a_0=0.2$ fm$^{-2}$.
With this potential we get a resonance energy of $E_R=197$ keV and a width
of the resonance $\Gamma\sim15$ keV. The difference in the width
compared to the value reported by Ref.~\cite{JH13} is mainly due to the
fact we do not include interferences with the ${}^3{\rm He}-\,^4{\rm He}$
channel.
Then we rewrite the total cross section as
\begin{equation}
  \sigma(E)={\cal{S}}_0^2\sigma^{\rm{bare}}_0(E)+
        {\cal{S}}^2_1\sigma^{\rm{bare}}_1(E)
        +{\cal{S}}_{\rm{res}}^2\sigma^{\rm{bare}}_{\rm{res}}(E)\,,
\label{eq:sigmae-he}
\end{equation}
where ${\cal{S}}_{\rm{res}}$ is the spectroscopic factor of the ${}^4P_{3/2}$
wave component in the GS and $\sigma^{\rm{bare}}_{\rm{res}}$ is
the calculated capture reaction cross section in the resonance wave.
The result obtained imposing ${\cal{S}}_0={\cal{S}}_1\sim1$ and
${\cal{S}}_{\rm{res}}\sim0.011$ is in good agreement
with both the data set of Refs.~\cite{ZS79} and~\cite{JH13}
and it is shown in Fig.~\ref{fig:sfres}. 
The small value of ${\cal{S}}_{\rm{res}}$ reflects
the small percentage of spin 3/2 component in the $\Be$ GS.
To be noticed that our results are also consistent with the
R-matrix fit reported in Ref.~\cite{JH13}. 
However, using the potential model which describes the resonance
in the $S$-factor data, we
were not able to reproduce the ${}^4S_{3/2}$ elastic phase shifts data.
In fact, as shown in Fig.~\ref{fig:phsres}, the ${}^4S_{3/2}$ phase shift
is badly underpredicted.
Therefore we can conclude that by including the resonance structure
in the  ${}^4S_{3/2}$ wave, we obtain a nice description of the $S$-factor
data, but we destroy the agreement between theory and experiment for the
elastic phase shifts. This put under question the
real existence of the resonance structure proposed in Ref.~\cite{JH13}.

\section{Summary and Conclusions} 
\label{sec:concl} 

We have evaluated the astrophysical
$S$-factor of the $\Li(p,\gamma)\Be$
radiative capture reaction using a two-body cluster approach.
The intercluster potential parameters are fitted to reproduce
the bound state properties of $\Be$ and the scattering phase shifts for
all the partial waves of interest.
The wave functions are calculated solving the 
Schr\"odinger equation with the Numerov method.

The theoretical $S$-factor underestimates,
although not dramatically, the experimental values.
This is not surprising, since we have neglected in this first step
the internal structure
of the involved nuclei. When we introduce a spectroscopic factor,
which takes care of this, we obtain a nice agreement with the data.
Furthermore, we have reviewed the phenomenological calculations present
in literature, and this has allowed us to estimate the
theoretical error on the $S$-factor calculated within this two-body
approach.

We have also studied the photon angular distribution. Comparing our calculations
with the available data, we obtain a good description
of the angular distribution for the capture to the GS of $\Be$.
For the capture to the FES, the description is less accurate,
and this can be traced back to the poor knowledge
of the $P$ and $D$ waves phase shifts.
Furthermore, we have used our calculation of the photon angular
distribution to study the effect on the experimental error budget
for those experiments which measure the $\Li(p,\gamma)\Be$
cross section using a fixed angle apparatus.

Finally, we have introduced in our study the resonance-like structure
proposed in Ref.~\cite{JH13}. If the resonance is introduced
in the ${}^2S_{1/2}$ or in the ${}^2D_{3/2}$ waves, we obtain results
completely inconsistent with the phase shifts and S-factor data.
When
the resonance is introduced in the ${}^4S_{3/2}$ wave,
a nice description of the $S$-factor data is achieved.
However, we are not able to reproduce consistently
the ${}^4S_{3/2}$ elastic scattering phase shifts.
We can conclude therefore that the presence of
a resonant structure cannot be accepted in our theoretical
framework.


\section*{Acknowledgement}
The Authors are grateful to the LUNA Collaboration, and especially
R.\ Depalo, L.\ Csedreki, and G.\ Imbriani, for comments and useful
discussions. The Authors acknowledge useful discussions with R.J. deBoer,
who suggested to perform the theoretical investigation of the angular distribution.



\end{document}